\documentclass{elsart}

\usepackage{harvard}

\usepackage{epsfig}

\usepackage{amssymb}

\def\url#1{{\ttfamily\def\/{/\discretionary{}{}{}}#1}}

\begin{document}

\begin{frontmatter}
\title{Accretion onto a primordial protostar}
\author{Volker Bromm},
\ead{vbromm@cfa.harvard.edu}
\author{Abraham Loeb}
\ead{aloeb@cfa.harvard.edu}
\address{Department of Astronomy, Harvard University, Cambridge, MA 02138, USA}

\begin{abstract}
We present a three-dimensional numerical simulation that resolves the
formation process of a Population~III star down to a scale of $\sim 100$
AU. The simulation is initialized on the scale of a dark matter halo of
mass $\sim 10^6 M_{\odot}$ that virializes at $z\sim 20$. It then follows
the formation of a fully-molecular central core, and traces the accretion
from the diffuse dust-free cloud onto the protostellar core for as long as
$\sim 10^4$~yr, at which time the core has grown to $\sim 50 M_{\odot}$.
We find that the accretion rate starts very high, $\sim 0.1
M_{\odot}\mbox{\,yr}^{-1}$, and declines rapidly thereafter approaching a
power-law temporal scaling. Asymptotically, at times $t\gtrsim 10^3
\mbox{\,yr}$ after core formation, the stellar mass grows approximately as
$M_{\ast}\simeq 20 M_{\odot}(t/10^3\mbox{\,yr})^{0.4}$.  Earlier on,
accretion is faster with $M_{\ast}\propto t^{0.75}$.  By extrapolating this
growth over the full lifetime of very massive stars, $t\simeq 3\times
10^6$~yr, we obtain the conservative upper limit $M_{\ast}\lesssim 500
M_{\odot}$. The actual stellar mass is, however, likely to be significantly
smaller than this mass limit due to radiative and mechanical feedback from
the protostar.
\end{abstract}

\end{frontmatter}

\section{Introduction}
\label{intro}

The first (so-called Population~III) stars could have had a dramatic
influence on the early universe at redshifts $z\lesssim 20$. Their high
yield of ionizing photons may have reionized the intergalactic medium (IGM)
(e.g., Cen, 2003; Ciardi et al., 2003; Haiman and Holder, 2003; Sokasian et
al., 2003; Somerville and Livio, 2003; Wyithe and Loeb, 2003).  In
addition, the first supernovae (SNe) were responsible for the initial IGM
metal-enrichment (e.g., Madau et al., 2001; Mori et al., 2002; Schneider et
al., 2002; Bromm et al., 2003; Furlanetto and Loeb, 2003; Mackey et al.,
2003; Scannapieco et al., 2003; Wada and Venkatesan, 2003; Norman et al.,
2004; Yoshida et al., 2004).  The radiative and hydrodynamic feedback of
the first stars affected subsequent galaxy formation (Barkana and Loeb,
2001) and imprinted large-scale polarization anisotropies on the cosmic
microwave background (Kaplinghat et al., 2003; Kogut et al., 2003).  In the
context of popular cold dark matter (CDM) models of hierarchical structure
formation, the first stars are predicted to have formed in dark matter
halos of mass $\sim 10^{6}M_{\odot}$ that collapsed at redshifts $z\simeq
20-30$ (e.g., Tegmark et al., 1997; Barkana and Loeb, 2001; Yoshida et al.,
2003). Since their ionization efficiency (Tumlinson and Shull, 2000; Bromm
et al., 2001; Schaerer, 2002) and metal yield (Heger and Woosley, 2002;
Umeda and Nomoto, 2002, 2003) depends on their mass, 
the fundamental question is (e.g., Bromm, 2004):
{\it How massive were the first stars?}

Results from recent numerical simulations of the collapse and fragmentation
of primordial clouds suggest that the first stars were predominantly very
massive, with typical masses $M_{\ast}\gtrsim 100 M_{\odot}$ (Bromm et al.,
1999, 2002; Nakamura and Umemura, 2001; Abel et al., 2002). More
specifically, these simulations have identified pre-stellar clumps of
masses up to $\sim 10^{3}M_{\odot}$ and sizes $\lesssim 0.5$~pc.  Each of
the simulated clumps is conjectured to be the immediate progenitor of a
single star or a small cluster of stars (see Larson 2003 for a review). To
determine the mass of a single star, one must follow the fate of such a
clump.  Extending the analogous calculation for the collapse of a
present-day protostar (Larson, 1969) to the primordial case, Omukai and
Nishi (1998) have carried out one-dimensional hydrodynamical simulations in
spherical symmetry. They have found that the mass of the initial
hydrostatic core, formed near the center of the collapsing cloud when the
density is sufficiently high ($n\sim 10^{22}$~cm$^{-3}$) for the gas to
become optically thick to continuum radiation, is almost the same as in
present-day star formation: $M_{\rm core}\sim 5\times
10^{-3}M_{\odot}$. The small value of the initial core has no bearing on
the final mass, which is in turn determined by how efficient the accretion
process is at incorporating the clump mass into the growing protostar.

In this paper, we present results from the first three-dimensional
simulation that is initialized on cosmological scales, followed to the
formation of a protostellar high-density core, and for which the accretion
flow is traced onto the central protostar for as long as $\sim 10^4$~yr
after core formation. We do not yet self-consistently take into account the
radiative and mechanical feedback from the protostar on the accretion flow
(Omukai and Palla, 2001, 2003; Tan and McKee, 2003).  By comparing to
one-dimensional calculations that address this feedback on the accretion
flow, we demonstrate that the accretion process is likely to lead to the
formation of large stellar masses, as conjectured before.

Throughout this paper, we assume a standard $\Lambda$CDM cosmology, with a
total density parameter in matter of $\Omega_{m}=1- \Omega_{\Lambda}=0.3$,
and in baryons of $\Omega_{\rm B}=0.045$, a Hubble constant $h=H_{0}/(100$
km s$^{-1}$ Mpc$^{-1})=0.7$, and a power-spectrum amplitude
$\sigma_{8}=0.9$ on $8h^{-1}$Mpc spheres.  These values reflect the latest
estimates of cosmological parameters from the {\it Wilkinson Microwave
Anisotropy Probe (WMAP)} (Spergel et al., 2003).

\section{Physical ingredients}

The main physical processes operating prior to the formation of a
pre-stellar clump with typical mass of a few hundred solar masses are
described in earlier work (Bromm et al., 2002; hereafter BCL).  Below we
discuss the physical effects that become important during the subsequent
collapse to higher densities and the resultant formation of a protostellar
core (see also Bromm, 2000).

The formation of hydrogen molecules due to the H$^{-}$ channel saturates at
a fractional abundance of $\sim 10^{-3}$ (BCL).  At densities exceeding
$\sim 10^8$~cm$^{-3}$, however, three-body reactions are able to convert
the gas into an almost fully molecular form (Palla et al., 1983). In
addition to the reactions discussed by BCL (see their Table~1), we also
consider here
\begin{displaymath}
(13)\mbox{\ \ \ \ }  \mbox{H}+ \mbox{H} +\mbox{H} \rightarrow \mbox{H}_{2} + \mbox{H}
\end{displaymath}
\begin{displaymath}
(14)\mbox{\ \ \ \ }  \mbox{H}+ \mbox{H} +\mbox{H}_{2} \rightarrow \mbox{H}_{2} + \mbox{H}_{2}
\end{displaymath}
The reaction rates are (Palla et al., 1983): $k_{13}=5.5\times
10^{-29}(T/{\rm K})^{-1}$ cm$^{6}$ s$^{-1}$, and $k_{14}=\frac{1}{8}
k_{13}$.  To estimate the density at which three-body reactions become
important, we consider reaction (13)
\begin{equation}
\frac{\mbox{d}n{\mbox{\scriptsize [H$_{2}$]}}}{\mbox{d}t}=
k_{13}\left(n\mbox{\scriptsize [H]}\right)^{3}
\mbox{\ \ \ .}
\end{equation}
Defining the fractional abundance of H$_{2}$ molecules as
$f=2n\mbox{\scriptsize [H$_{2}$]}/n_{\mbox{\scriptsize H}}$, and expressing
the density of hydrogen atoms $n\mbox{\scriptsize [H]}=(1-f)
n_{\mbox{\scriptsize H}}$ with $f\sim 10^{-3} \ll 1$, we get
the timescale for reaction (13)
\begin{equation}
t_{\mbox{\scriptsize 3-body}}\simeq \frac{f}{2 k_{13} n^{2}_{\rm H}} .
\end{equation}
The value of $t_{\mbox{\scriptsize 3-body}}$ equals the free-fall timescale
at a critical density
\begin{equation}
n_{\rm H}\simeq \left( \frac{f^{2} G m_{\rm H}}{4 k_{13}^{2}}\right)^{1/3}
\mbox{\ \ \ .}
\end{equation}
Evaluating $k_{13}$ at $T\simeq 10^3$ K, and assuming $f\simeq 10^{-3}$, we
find that three-body reactions set in at $n_{\rm H}\simeq 2\times 10^{8}$
cm$^{-3}$. Along similar lines, by considering reaction (14) and assuming
$f\simeq 0.5$, we find that the density at which the conversion of hydrogen
atoms into molecules is complete is $n_{\rm H}\simeq 5\times 10^{11}$
cm$^{-3}$.  Both estimates are in good agreement with our numerical
results. The transformation of the hydrogen gas from an atomic phase to a
molecular phase requires three modifications of the physics incorporated in
the numerical code as described below (see also Omukai and Nishi, 1998).

First, the heating associated with the formation of H$_{2}$
and the corresponding release of the molecular binding 
energy, $\epsilon_{\rm H_{2}}=4.48$~eV, becomes important at high densities.
To account for this additional heating source, we add to
the energy equation (see BCL) the term (in ergs s$^{-1}$ cm$^{-3}$) 
\begin{equation}
\Gamma_{\mbox{\scriptsize 3-body}}=\epsilon_{\rm H_{2}}
\frac{\mbox{d}n{\mbox{\scriptsize [H$_{2}$]}}}{\mbox{d}t}
\mbox{\ \ \ ,}
\end{equation}
for densities $n_{\rm H}\ge 10^{8}$~cm$^{-3}$.

Second, the H$_{2}$ cooling function has to be augmented to include
both the collisional excitation of H$_{2}$ by H atoms and by H$_{2}$
molecules.  The total cooling rate (in ergs s$^{-1}$ cm$^{-3}$) can then be
written as
\begin{equation}
\Lambda= n^{2}_{\rm H}f\left[
\frac{n\mbox{\scriptsize [H]}}{n_{\rm H}}L^{\rm H} + f L^{\rm H_2}
\right]
\mbox{\ \ \ ,}
\end{equation}
where $L^{\rm H}$ and $L^{\rm H_2}$ (in erg s$^{-1}$ cm$^{3}$) are the
vibrational/rotational cooling coefficients for collisions with H atoms and
H$_{2}$ molecules, respectively.  At the high densities considered here,
all levels are close to being populated according to local thermodynamic
equilibrium (LTE). For $L^{\rm H}$ and $L^{\rm H_2}$, we use the
parameterization given by Hollenbach and McKee (1979).

Third, the presence of molecules leads to a modified equation of state.
We write the gas pressure as
\begin{equation}
P= \frac{k_{\rm B} T}{\mu m_{\rm H}}\rho=(\gamma_{\rm ad}-1)\rho u
\mbox{\ \ \ ,}
\end{equation}
where $\gamma_{\rm ad}$ is the adiabatic exponent, and $u$ is the specific
internal energy. The mean molecular weight, $\mu$, is given by
\begin{equation}
\frac{1}{\mu}=\frac{Y}{4}+(1-f)X+f\frac{X}{2}
\mbox{\ \ \ ,}
\end{equation}
where $X=0.76$ and $Y=1-X$ are the mass fractions in hydrogen and helium,
respectively. For a gas consisting of helium and pure atomic hydrogen,
$\mu\simeq 1.22$, whereas for a mixture of helium and pure
molecular hydrogen, $\mu\simeq 2.27$.
The adiabatic exponent can be expressed as
\begin{equation}
\gamma_{\rm ad}-1=\frac{1}{\mu}\left(\sum_{i}\frac{1}{(\gamma_{i}-1)
\mu_{i}}\right)^{-1}
\mbox{\ \ \ ,}
\end{equation}
where the summation is over the 6 species
H, H$^{+}$, H$^{-}$, H$_{2}$, e$^{-}$, and He, with $\mu_{\rm H_2}=2$,
$\mu_{\rm He}=4$, and $\mu_{i}=1$ otherwise.  In calculating the adiabatic
exponent for H$_{2}$, one has to consider translational, rotational, and
vibrational degrees of freedom, resulting in
\begin{equation}
\frac{1}{\gamma_{\rm H_{2}}-1}=\frac{1}{2}\left(
3+2+\frac{2\bar{\epsilon}_{\rm vib}}{k_{\rm B} T}
\right)
\mbox{\ \ \ .}
\end{equation}
Here, $\bar{\epsilon}_{\rm vib}$ is the average vibrational energy per
H$_2$ molecule, and is given by (e.g., Shu, 1991)
\begin{equation}
\bar{\epsilon}_{\rm vib}=\frac{\hbar\omega_{0}}{{\rm e}^{ \hbar\omega_{0}/
k_{\rm B} T} -1} + \frac{1}{2}\hbar\omega_{0}
\mbox{\ \ \ ,}
\end{equation}
with $\hbar\omega_{0}/k_{\rm B}\simeq 6300$ K.  For the physical conditions
considered here, the excitation of vibrational degrees of freedom is never
important. Other non-molecular species have only translational degrees of
freedom, and consequently, $1/(\gamma_{i}-1)=3/2$.  We terminate the
simulation when the central density is sufficiently high for opacity
effects to be important. In the following, we estimate analytically the
density beyond which the H$_{2}$ line radiation cannot escape from the
central fully-molecular core of the gas.

Molecular opacity starts to affect the cooling rate of the gas at a
sufficiently high density.  At temperatures $T\sim 10^3$ K, only the
rotational transitions within the lowest-lying vibrational level contribute
significantly to the cooling. Taking into account the quadrupole nature of
the transitions, corresponding to the selection rule $\Delta J=2$ (where
$J$ is the angular momentum quantum number), one may express the H$_{2}$
cooling function as
\begin{equation}
\Lambda=\sum_{J\rightarrow J-2} n_{J}A_{J,J-2}\Delta E_{J,J-2}
\mbox{\ \ \ .}
\end{equation}
The number density of molecules in rotational level $J$ is given by
\begin{equation}
n_{J}=\frac{n\mbox{\scriptsize [H$_2$]}}{Z}(2J+1)\mbox{e}^{-E_{J}/k_{\rm B}T}
\mbox{\ \ \ ,}
\end{equation}
where $n\mbox{\scriptsize [H$_2$]}$ is the total H$_{2}$ number density,
$E_{J}=85.3 \mbox{\,K\ }J(J+1)$ is the energy corresponding to level $J$,
and $Z$ is the partition function 
\begin{equation}
Z=\sum_{J=0}^{\infty} (2J+1)\mbox{e}^{-E_{J}/k_{\rm B}T}
\mbox{\ \ \ .}
\end{equation}
The energy difference between levels $J$ and $J-2$ is $\Delta E_{J,J-2}$,
and the corresponding probability per unit time for spontaneous emission is
$A_{J,J-2}\propto \Delta E_{J,J-2}^{5}$. We first identify the transition
that dominates the radiative cooling at a given temperature,
and then determine the opacity in this most important line. The most 
important transition is near the extremum set by the condition
\begin{equation}
\frac{\mbox{d}}{\mbox{d}J}(n_{J}A_{J,J-2}\Delta E_{J,J-2})=0
\mbox{\ \ \ ,}
\end{equation}
implying
\begin{equation}
J_{\rm max}\simeq 6.4\left(\frac{T}{\mbox{1000\,K}}\right)^{1/2}
\mbox{\ \ \ ,}
\end{equation}
where $T\sim 10^3$ K is typical for the central region of the collapsing
clump.

For the $6\rightarrow 4$ transition, the line absorption coefficient (in
cm$^{-1}$) can be expressed as (Rybicki and Lightman, 1979)
\begin{equation}
\alpha_{\nu}=\frac{\Delta E_{6,4}}{4\pi}n_{4}B_{4,6}\left[
1-\mbox{e}^{-\Delta E_{6,4}/k_{\rm B}T}
\right]\varphi(\nu)
\mbox{\ \ \ ,}
\end{equation}
where we have taken the levels to be populated according to LTE.
By assuming that broadening is entirely due to the thermal Doppler effect,
the profile function is
\begin{equation}
\varphi(\nu)=\frac{1}{\sqrt{\pi}\Delta \nu_{D}}\exp\left[-\frac{(\nu -\nu_{0})^{2}}
{\Delta \nu_{D}^{2}}\right]
\mbox{\ \ \ ,}
\end{equation}
with $\Delta \nu_{D}=(\nu_{0}/c)\sqrt{2 k_{\rm B} T/m}$, a particle mass
$m=2 m_{\rm H}$ for H$_{2}$, and $h\nu_{0}=\Delta E_{6,4}$.  This function
leads to a modest overestimate of the opacity, as parts of the accretion flow
are mildly
supersonic (with a maximum Mach number of $\sim 1.5$ at radius $\sim 200$~AU), 
but our goal here is only to get a rough estimate as to when line opacity
becomes important.  In the following, we approximate $\varphi(\nu)\sim
\varphi(\nu_{0})= 1/(\sqrt{\pi}\Delta \nu_{D})$, and use $\Delta
\nu_{D}/\nu_{0}\simeq 10^{-5}$ for $T\sim 10^{3}$ K. The Einstein
coefficient for stimulated absorption is
\begin{equation}
B_{4,6}=\frac{13}{9}\frac{c^{2}h^{2}}{2(\Delta E_{6,4})^{3}}A_{6,4}\simeq 0.04
{\rm \ cm}^2{\rm \  erg}^{-1}{\rm \  s}^{-1}\mbox{\ \ ,}
\end{equation}
where $A_{6,4}$ is taken from Turner et al. (1977).  At $T\sim 10^{3}$ K,
the partition function is $Z\simeq 12.1$, and using equation (15) we find
$n_{4}\simeq 0.1 n\mbox{\scriptsize [H$_2$]}$.

We may now estimate the density at which the optical depth,
$\tau_{\nu}\simeq \alpha_{\nu}L$, approaches unity, where $L$ is the
characteristic size of the high-density, fully molecular region
\begin{equation}
n_{\rm H}\simeq 3\times 10^{6}
\mbox{\,cm$^{-3}$\,}\left(\frac{L}{\mbox{pc}}\right)^{-1} \mbox{\ \ \ .}
\end{equation}
From our simulation, we estimate $L\simeq 10^{-4}$ pc, and consequently
have $n_{\rm H}\simeq 10^{11}$ cm$^{-3}$. At densities exceeding this
value, radiation can still escape in less opaque lines with a smaller $A$
coefficient, and in the continuum between the lines, but neglecting the
effects of radiative transfer renders the results of the simulation
increasingly unreliable at yet higher densities.  Omukai and Nishi (1998),
who did include the treatment of radiative transfer, were able to follow
the collapse all the way to stellar densities, although their approach was
limited to a spherical geometry.
In our numerical simulation, we form a sink particle once the
gas density exceeds a value of $10^{12}$~cm$^{-3}$ (see BCL for details
about the numerical implementation).  This sink particle approximates a
hydrostatic core at the center of the accretion flow, although it has a
size that is still much larger than the expected scale of the actual core
(Omukai and Nishi, 1998; Ripamonti et al., 2002).

\section{Numerical methodology}

The evolution of the dark matter and gas components is calculated with a
version of TREESPH (Hernquist and Katz, 1989), combining the smoothed
particle hydrodynamics (SPH) method with a hierarchical (tree) gravity
solver (see BCL for further details).  Here, we briefly describe the
modifications introduced to the code in order to allow it to treat
accretion flows in a dense, high-redshift environment.

We initialize the simulation (see \S~4.1) on the scale of the minihalo
($\gtrsim 150$~pc), and follow the collapse to the point where the gas is
converted into a fully molecular phase (on a scale $\lesssim 100$~AU).  In
order to treat the large dynamic range spanned between the cosmological and
protostellar length scales, we carry out our simulation in two steps. We
first simulate the collapse of the primordial gas, together with the
virialization of the dark matter (DM) in the minihalo, until the formation of a
gravitationally-bound clump (similar to BCL). At this point, we stop the
simulation, and resample the central density field (involving $\sim 3000
M_{\odot}$ in gas within a spherical region of radius $3.1$~pc), with an
increased number of SPH particles.  We adopt a rapidly decreasing timestep
according to the scaling $\Delta t_{\rm sys}\propto 1/\sqrt{G\rho_{\rm
max}}$.  Here, $\rho_{\rm max}$ is the maximum gas density at a given
instant and the system timestep, $\Delta t_{\rm sys}$, is the maximum time
by which a particle is advanced within the multiple-timestep scheme
employed by the simulations (see Hernquist and Katz, 1989, for
details). Over the short timestep dictated by the maximum density, the
global system does not evolve by much, while the runaway collapse of the
compact dense core converges rapidly.

In general, the mass
resolution of an SPH simulation is approximately
\begin{equation}
M_{\rm res}=\left(\frac{1.5 N_{\rm neigh}}{N_{\rm SPH}}\right)M_{\rm B}
\mbox{\ \ \ ,}
\end{equation}
where $N_{\rm neigh}\simeq 32$ is the number of neighboring particles
within a given SPH smoothing kernel, $N_{\rm SPH}$ the total number of SPH
particles, and $M_{\rm B}$ the total baryonic mass. To avoid numerical
artifacts the Jeans mass, $M_J$, has to be resolved (Bate et al.,
2003), and so we require $M_{\rm res}<M_J$.  After the onset of
gravitational instability in the large-scale simulation, the rapid increase
in density naturally would lead to a violation of this criterion. The
resampling technique, however, allows us to follow the collapse to a much
higher density due to the improved $M_{\rm res}$.

We here briefly summarize the resampling procedure (see Bromm, 2000, and
Bromm and Loeb, 2003, for details).  Every SPH particle in the original,
unrefined simulation acts as a parent particle $p$, and spawns $N_{\rm
ref}$ child particles $i$, where $N_{\rm ref}=50$ in this paper. The child
particles are distributed according to the SPH smoothing kernel
$W(\vec{r}_{i}-\vec{r}_{p}; h_{p})$, by employing a standard Monte-Carlo
comparison-rejection method.  Here, $h_{p}$ is the smoothing length of the
parent particle. The velocity, temperature, H$_{2}$ abundance, and
free-electron abundance are directly inherited from the parent particle,
$\vec{v}_{i}=\vec{v}_{p}$, $T_{i}=T_{p}$, $f_{i}=f_{p}$, and $x_{i}=x_{p}$
for all $i$, respectively. Finally, each child particle is assigned a mass
$m_{i}=m_{p}/N_{\rm ref}$. This procedure guarantees the conservation of linear
and angular momentum. Energy is also conserved well, although there is a
small artificial contribution to the gravitational potential energy due to
the discreteness of the resampling.  The resampling results in $N_{\rm
SPH}=$ 35,500 within the central high-resolution region, and the mass
resolution is now $M_{\rm res}\simeq 4 M_{\odot}$, as compared to
$M_{\rm res}\simeq 200 M_{\odot}$ in the original simulation.  Our technique of
refining a coarser, parent simulation, and following the further collapse
with increased resolution, is conceptually similar to the adaptive mesh
refinement (AMR) method which 
has already been successfully applied to problems
in star formation (e.g., Truelove et al., 1998; Abel et al., 2002), but our
approach is one of the first attempts of implementing such a scheme within
SPH (see also Kitsionas and Whitworth, 2002; Bromm and Loeb, 2003).

To study accretion onto the central core, the diffuse gas in the envelope
must be followed for a sufficiently long time (roughly equal to the local
free-fall timescale). It is, therefore, necessary to introduce a central
sink particle at some stage in the evolution.  Otherwise, we would have to
adopt smaller and smaller timesteps to be able to follow the central
collapse, and with limited computer resources the surrounding envelope
would not have time to be accreted onto the core.  BCL prescribed SPH
particles to merge into more massive ones, provided they exceed a
pre-determined density threshold.  The sink particle technique allows us to
study the ongoing accretion process over many dynamical times.

\section{Simulating the accretion flow}
\subsection{Initial Conditions}

Within the hierarchical $\Lambda$CDM model, the first luminous objects are
expected to form out of high-$\sigma$ peaks in the random field of
primordial density fluctuations.  The early (linear) evolution of such a
peak, assumed to be isolated and roughly spherical, can be described by the
top-hat model (e.g., Padmanabhan, 1993). We use the top-hat approximation
to specify the initial amplitude of the CDM configuration.  In this paper,
we investigate the fate of a high-$\sigma$ peak of total mass
$10^{6}M_{\odot}$, corresponding to $1.5\times 10^{5}M_{\odot}$ in baryons,
which collapses (or virializes) at a redshift $z_{\rm vir}\simeq 20$.

Our simulation is initialized at $z_{i}=100$, by performing the following
steps.  The collisionless DM particles are placed on a cubical Cartesian
grid, and are then perturbed according to a given power spectrum $P(k)=A
k^{n}$, by applying the Zeldovich (1970) approximation which also allows
to self-consistently assign initial velocities. The power-law index is set
to $n=-3$ which approximately describes the spectral behavior on the mass
scale of $\sim 10^{6}M_{\odot}$.  To fix the amplitude $A$, we specify the
initial variance of the fluctuations
\begin{equation}
\sigma_{i}^{2}=A\sum k^{n}
\mbox{\ \ \ .}
\end{equation}
The summation is over all contributing modes, where the minimum wavenumber
is given by the overall size of the Cartesian box, and the maximum
wavenumber is set by the Nyquist frequency, $k_{\rm max}\simeq {2\pi}/{d_{\rm
grid}} $. Here, $d_{\rm grid}$ is the mesh size of the Cartesian grid.
Choosing $\sigma_{i}\simeq 0.2$ at $z_i=100$, the rms fluctuation at the
moment of collapse is
$\sigma(z=20)=[(1+z_{i})/(1+z)]\sigma_{i}\simeq 1$.  This choice
ensures that substructure develops on a similar timescale as the overall
collapse of the background medium.  Next, particles within a (proper)
radius of $R_{i}\simeq 160$~pc are selected for the simulation. The
resulting number of DM particles is $N_{\rm DM}=17,074$.  Finally, the
particles are set into rigid rotation and are endowed with a uniform Hubble
expansion (see Katz, 1991).  Angular momentum is added by assuming a spin
parameter $\lambda_s=L|E|^{1/2}/(G M^{5/2}) = 0.05$, close to the
mean-value measured in cosmological N-body simulations (e.g., Jang-Condell
and Hernquist, 2001).  Here, $L$, $E$, and $M$ are the total angular
momentum, binding energy, and mass, respectively. The spin parameter is a measure
of the degree of rotational support, such that the ratio of centrifugal to
gravitational acceleration is given by $\sim \lambda_s^{2}$ at
virialization.  The collisional SPH particles ($N_{\rm SPH}=32,768$) are
randomly placed to approximate a uniform initial density.  The SPH
particles follow the same Hubble expansion and rigid rotation as the DM
particles.  

\subsection{Results}

The initial evolution up to the formation of a pre-stellar clump is very
similar to the cases studied by BCL. The gas dissipatively settles into the
center of the CDM minihalo, reaching the characteristic state with density
$\sim 10^{4}$~cm$^{-3}$ and temperature $\sim 200$~K.  In Figure~1 ({\it
left panel}), we show the morphology within the central $\sim 25$~pc,
briefly after the first high-density core has formed as a result of
gravitational runaway collapse.

\begin{figure}[t]
\epsfig{file=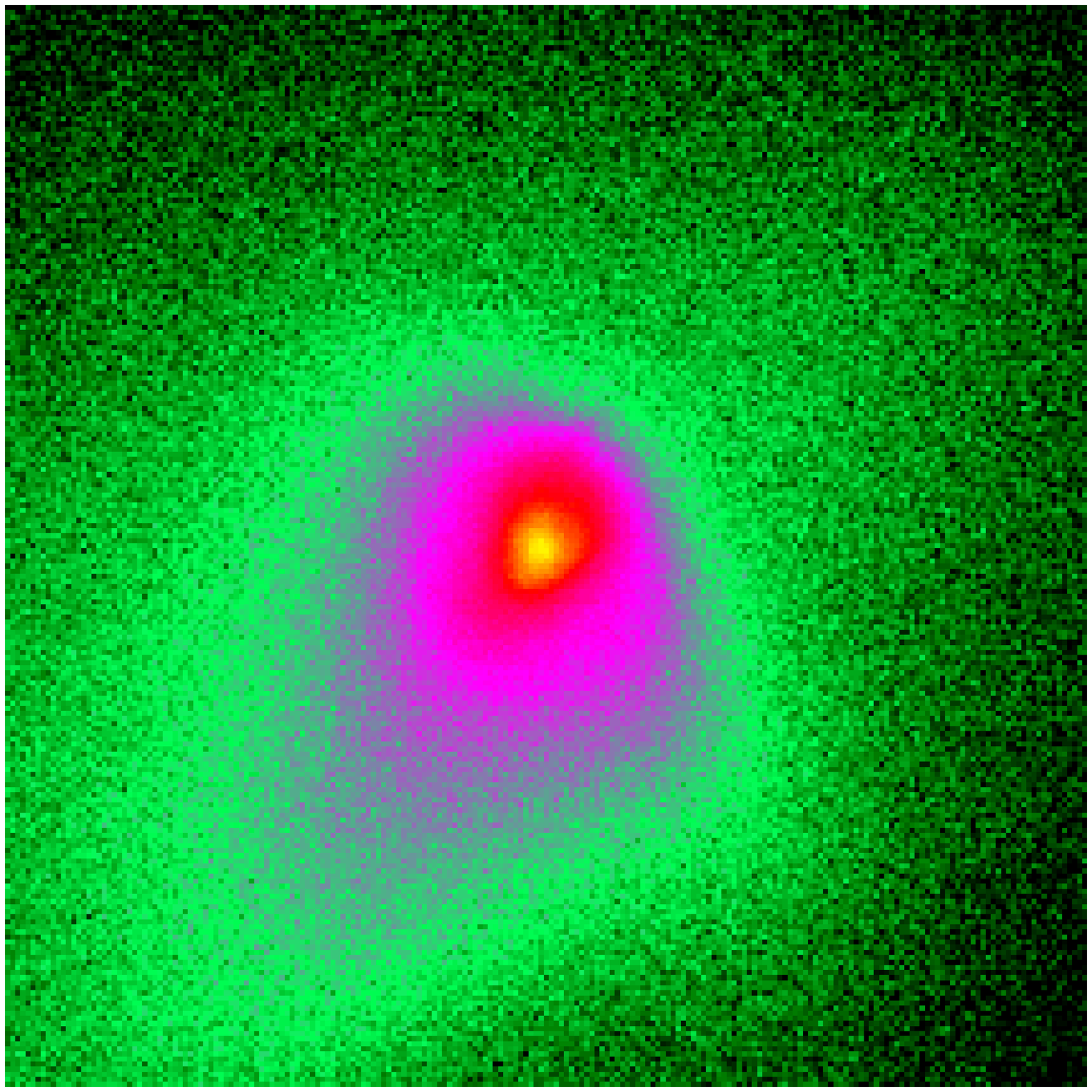,width=6.93cm,height=6.237cm}
\epsfig{file=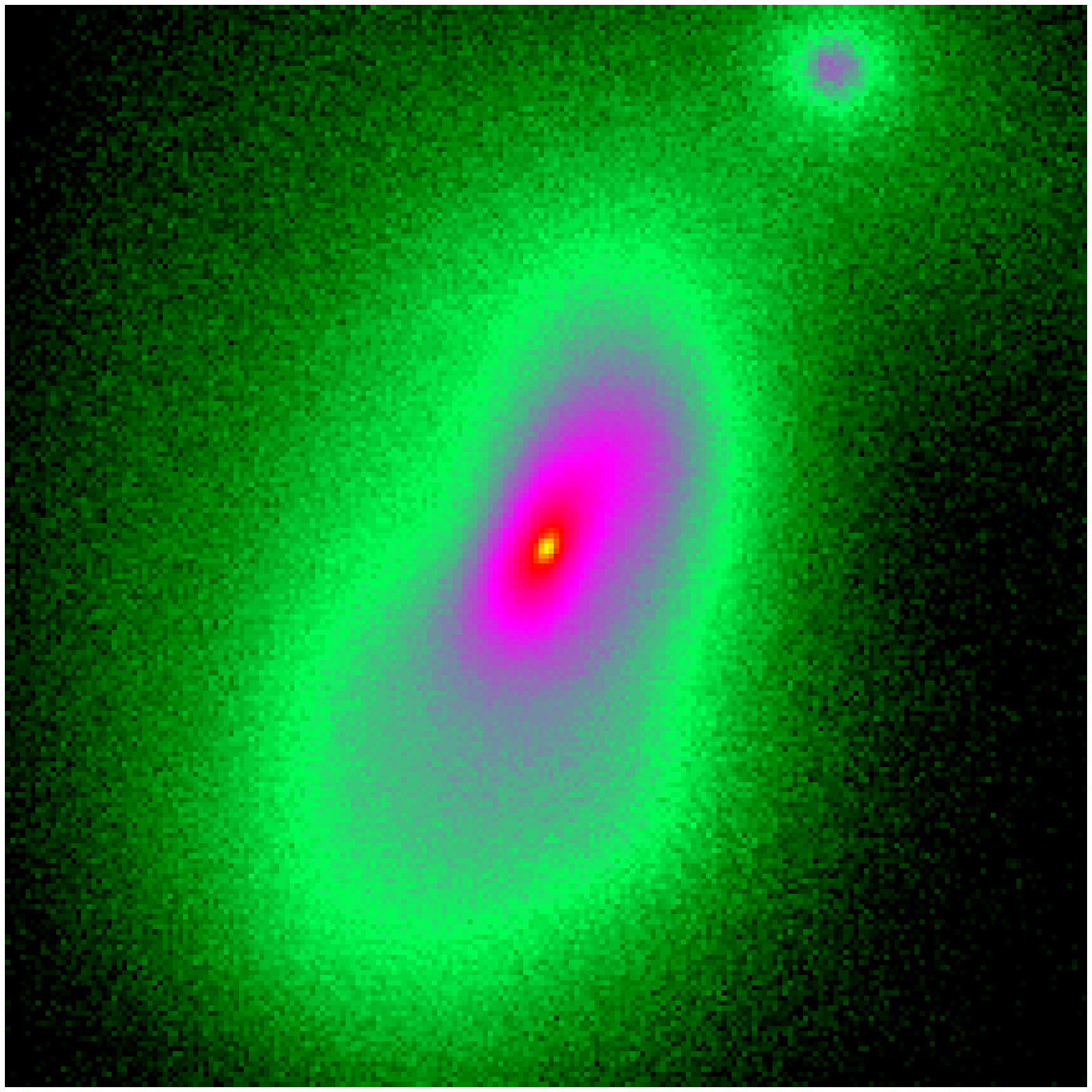,width=6.93cm,height=6.237cm}
\caption{Collapse and fragmentation of a primordial cloud.  We show the
projected gas density at a redshift $z\simeq 21.5$, shortly after
gravitational runaway collapse has converged at the center of the cloud.
{\it Left:} The coarse-grained morphology in a box with a linear physical
size of 23.5~pc.  At this time, a gravitationally-bound clump has formed
with a mass of $\sim 10^{3}M_{\odot}$.  {\it Right:} The fine-grained
morphology in a box with a linear physical size of 0.5~pc.  The central
density peak accretes mass vigorously, and is accompanied by a secondary
clump.}
\label{fig1}
\end{figure}

\begin{figure}[t]
\epsfig{file=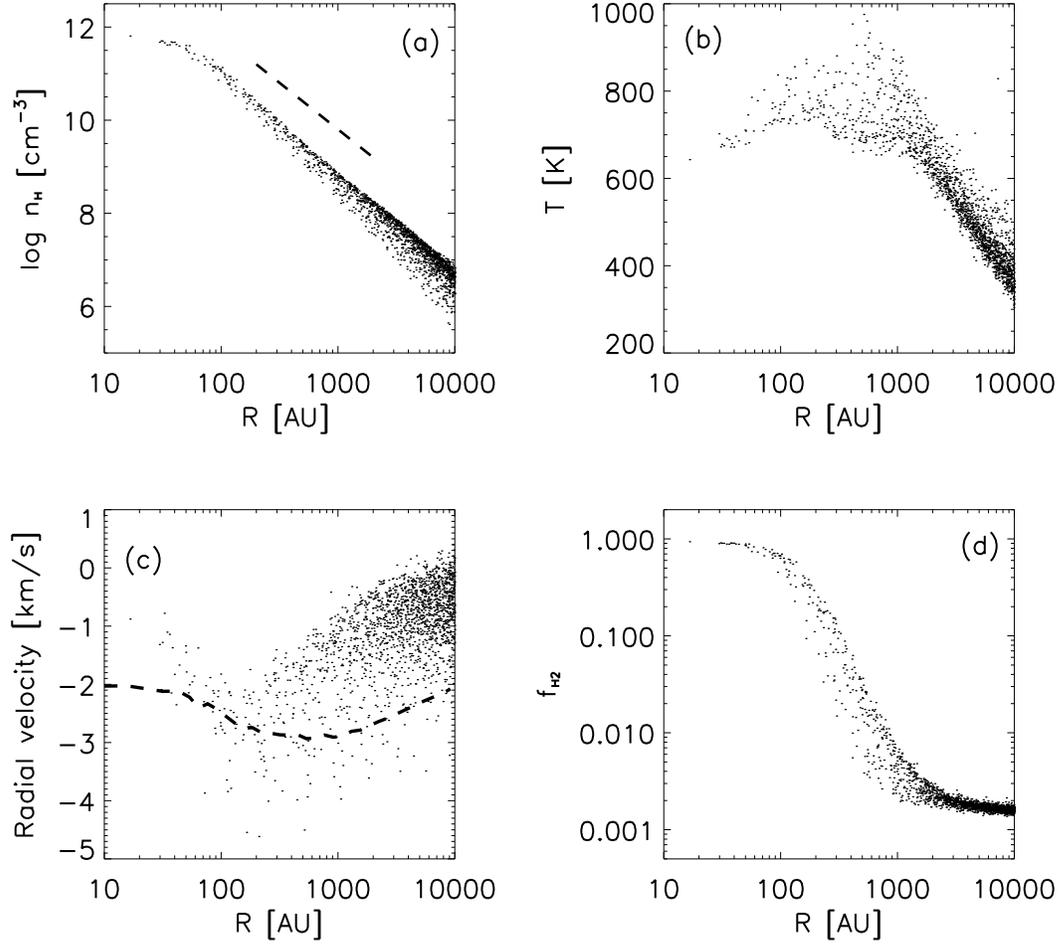,width=13.86cm,height=12.474cm}
\caption{Radial distribution of physical properties of SPH particles within
the main clump.  The presented snapshot was taken shortly before the
central sink particle was created.  {\bf (a)} Hydrogen number density (in
cm$^{-3}$) vs. radial distance from density maximum (in AU).  For
comparison, we also show the isothermal profile {\it (dashed line)}, which
provides a good description for the outer region. {\bf (b)} Gas temperature
(in K).  {\bf (c)} Radial velocity (in km s$^{-1}$). {\it Dashed line:} 
Radially averaged (negative) sound speed. It is evident that the flow
is mostly subsonic, with a few fluid elements having mildly supersonic
speeds.
{\bf (d)} Hydrogen molecule
fraction.  Within the central $\sim 100$~AU, three-body reactions have
converted the gas into a fully molecular phase.  }
\label{fig2}
\end{figure}

To study the formation of a Population~III star, we have re-simulated the
evolution of the central clump with sufficient resolution to follow its
collapse up to a limiting density of $n\sim 10^{12}$~cm$^{-3}$, at which
point opacity effects become important (see \S 2), and a sink particle is
created. The right panel in Figure~1 shows the gas density on a scale of
0.5~pc.  Several features are evident in this image. First, the central
clump does not undergo further sub-fragmentation, and is likely to form a
single Population~III star. Second, a companion clump is visible at a
distance of $\sim 0.25$~pc. If negative feedback from the first-forming
star is ignored, this companion clump would undergo runaway collapse on its
own, approximately $\sim 3$~Myr later.  This timescale is comparable to the
lifetime of a very massive star (e.g., Bromm et al., 2001).  If the second clump
were able to survive the intense radiative heating from its neighbor, it
could become a star before the first one explodes as a SN. Whether more
than one star can form in a low-mass halo thus crucially depends on the
degree of synchronization of clump formation.  Finally, the
non-axisymmetric disturbance induced by the companion clump, as well as the
angular momentum stored in the orbital motion of the binary system, allow
the system to overcome the angular momentum barrier for the collapse of the
central clump (see Larson, 2002), and to possibly form a wide binary star
system.

In Figure~2, we consider the radial structure around the central density
maximum, shortly before the accreting sink particle is created.  The
overall behavior is similar to previous results (Omukai and Nishi, 1998;
Abel et al., 2002; Ripamonti et al., 2002). The density profile [panel (a)]
consists of a central, flat core, surrounded by a roughly isothermal
envelope. Such a configuration is generically predicted for collapsing star
forming clumps (see Larson, 2003), and can approximately be described by
the Larson-Penston similarity solution. Three-body reactions have succeeded
in converting the central $\gtrsim 1 M_{\odot}$ into fully molecular form
[panel (d)]. Notice that the central temperature never drops precipitously
due to the boost in H$_{2}$ cooling, even though H$_{2}$ is now more
abundant by a factor of $\sim 10^{3}$.  The opposite effects of
compressional heating and H$_{2}$ formation heating balance the enhanced
cooling rate, and lead to an almost isothermal collapse in the central
region.

Next, we address the accretion flow onto the central sink particle by
tracing its growth for $\sim 10^4$~yr after its initial formation.  This
accretion process will determine how massive the star finally gets.  We
begin with a brief discussion of the basic physics of the accretion, then
describe the results from earlier 1D calculations, and finally compare
these results to our simulation.
In general, star formation typically proceeds from the `inside-out' through
the accretion of gas onto a central hydrostatic core (e.g., Larson, 2003).
Even though the
initial mass of the hydrostatic core in the Population~III case is very
similar to that in the Population~I/II case, the accretion process is
expected to be rather different. On dimensional grounds, the accretion rate
simply scales as the sound speed cubed over Newton's constant (or
equivalently the ratio of the Jeans mass and the free-fall time):
$\dot{M}_{\rm acc}\sim c_s^3/G \propto T^{3/2}$. A comparison of the
temperatures in present-day star forming regions ($T\sim 10$~K) with those
in primordial ones ($T\sim 200-300$~K) already indicates a difference in
the accretion rate of more than two orders of magnitude.

Omukai and Palla (2001, 2003) extended earlier work by Stahler et
al. (1986) and investigated the spherical accretion problem in considerable
detail, going beyond the simple dimensional argument given above.  Their
computational technique approximates the time evolution by considering a
sequence of steady-state accretion flows onto a growing hydrostatic
core. Somewhat counterintuitively, these authors identified a critical
accretion rate, $\dot{M}_{\rm crit}\sim 4\times 10^{-3}
M_{\odot}$~yr$^{-1}$, beyond which the protostar cannot grow to masses much
in excess of a few tens of solar masses.  The critical rate is predicted to
be almost independent of mass (Omukai and Palla, 2003). For smaller rates,
however, the accretion is predicted to proceed all the way up to $\sim 600
M_{\odot}$, i.e., of order the host clump mass.  The physical origin of the
critical accretion rate is that for ongoing accretion onto the core, the
luminosity must not exceed the Eddington value, $L_{\rm EDD}$, if the inner
region of the gaseous core is ionized by the central star. In our
simulation, we do not resolve the evolution of the fully molecular core to
temperatures and densities high enough to completely ionize the gas.  Our
estimate of the accretion rate, evaluated at the density where the gas
becomes fully molecular, is likely to be applicable also at the higher
density where the gas becomes fully ionized.  This extrapolation assumes
that the inflowing gas will not be diverted into outflows before reaching
the central protostar. It will be important to test this assumption in
future work, taking into account the magneto-hydrodynamics of possible
bipolar jets.  We now ask whether the growing star is able to settle onto
the main sequence (MS), without experiencing a strong radiative feedback on
the accretion flow.  Before the onset of hydrogen burning, the luminosity
is approximately given by $L_{\rm tot}\sim L_{\rm acc}\simeq G
M_{\ast}\dot{M}_{\rm acc}/R_{\ast}$.  Here, we have ignored the
contribution to the overall luminosity from Kelvin-Helmholtz (KH)
contraction for simplicity, as we are only aiming at an order-of-magnitude
estimate.  By setting $L_{\rm acc}\simeq L_{\rm EDD}$ it follows that
\begin{equation}
\dot{M}_{\rm crit}\simeq \frac{L_{\rm EDD}R_{\ast}}{G M_{\ast}}
\sim 5\times 10^{-3} M_{\odot}{\rm ~yr}^{-1} \mbox{\ ,}
\end{equation}
where $R_{\ast}\sim 5 R_{\odot}$, a typical value for a Population~III MS
star (e.g., Bromm et al., 2001). This is the critical accretion rate that
must not be exceeded upon reaching the MS. If it is surpassed, radiation
pressure on free electrons will drive a strong wind, thus preventing
further accretion.

It is possible to start out with accretion rates that are significantly
larger than $\dot{M}_{\rm crit}$, because at earlier times the protostellar
radius is much larger, $R_{\ast}\gtrsim 100 R_{\odot}$, and it only
gradually shrinks to the MS value in the course of KH
contraction. During the KH phase, the star is in hydrostatic
equilibrium, but not yet in thermal equilibrium, which is only achieved on
the MS when contraction temporarily stops. The 
above condition can only determine whether accretion is capable of
incorporating most of the diffuse envelope into the star, or whether it is
shut-off by radiation pressure early on. The precise stellar mass in the
latter case can only be determined by a detailed simulation that takes both
radiative-transfer and the hydrodynamics into account.  Such a fully
self-consistent calculation has not yet been achieved in three dimensions,
although Omukai and Palla (2003), as well as Tan and McKee (2003), have
treated this demanding problem in zero- and one dimension.

Realistic flows are expected to produce a time-dependent accretion rate,
and so their outcome depends on whether the accretion rate declines rapidly
enough to avoid exceeding the Eddington luminosity at some stage during the
evolution.  The main caveat concerning the Omukai and Palla (2001, 2003)
results involves the geometry of the infall.  A three-dimensional
accretion flow of gas with angular momentum will deviate from spherical
symmetry, and instead form a flattened rotating configuration such as
a disk. In this case, most of the photons will likely escape along the
rotation axis (where the gas density and the corresponding opacity is low)
whereas mass may flow unimpeded along the disk plane (see Tan and McKee,
2003).

\begin{figure}
\epsfig{file=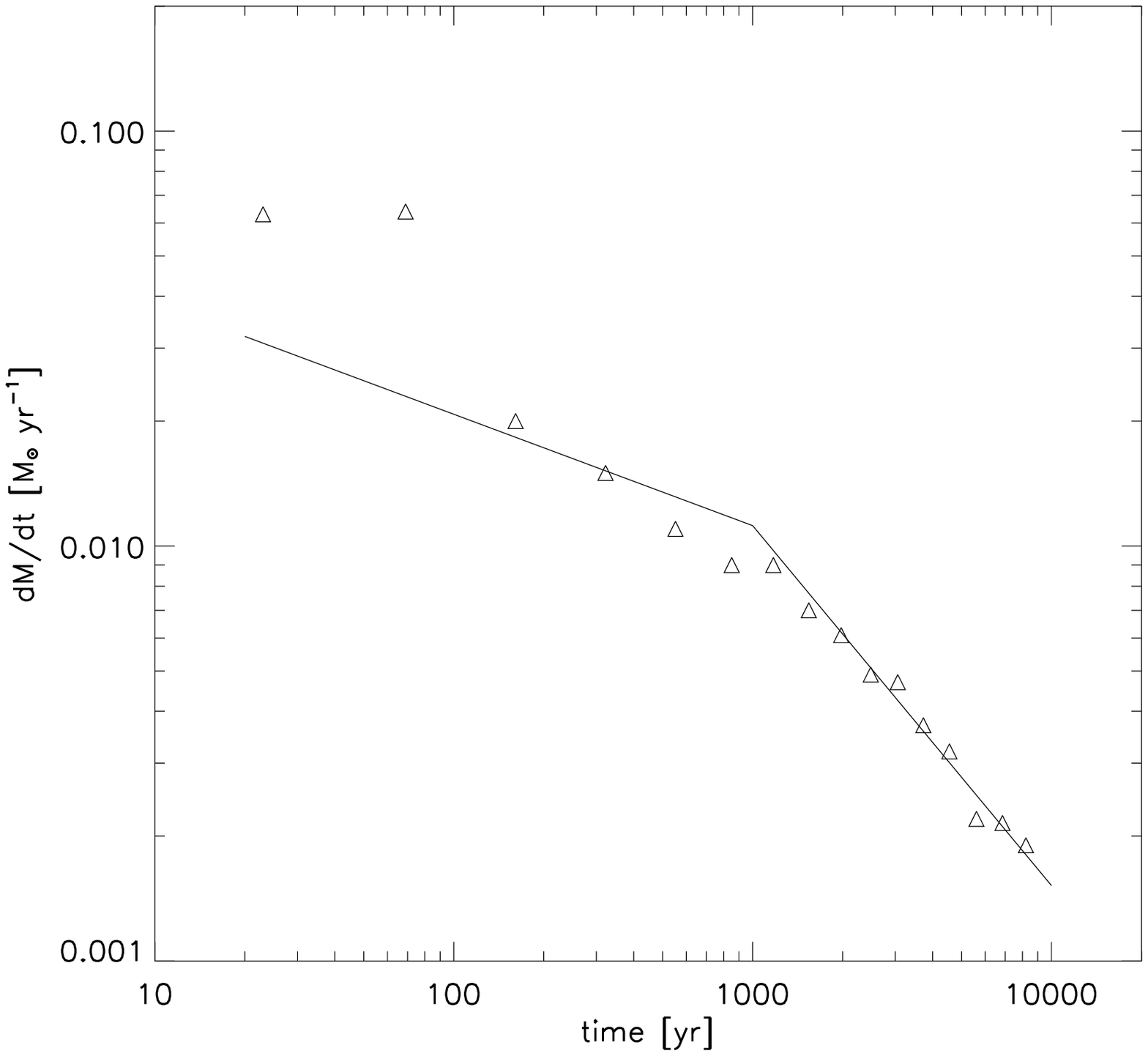,width=6.93cm,height=6.237cm}
\epsfig{file=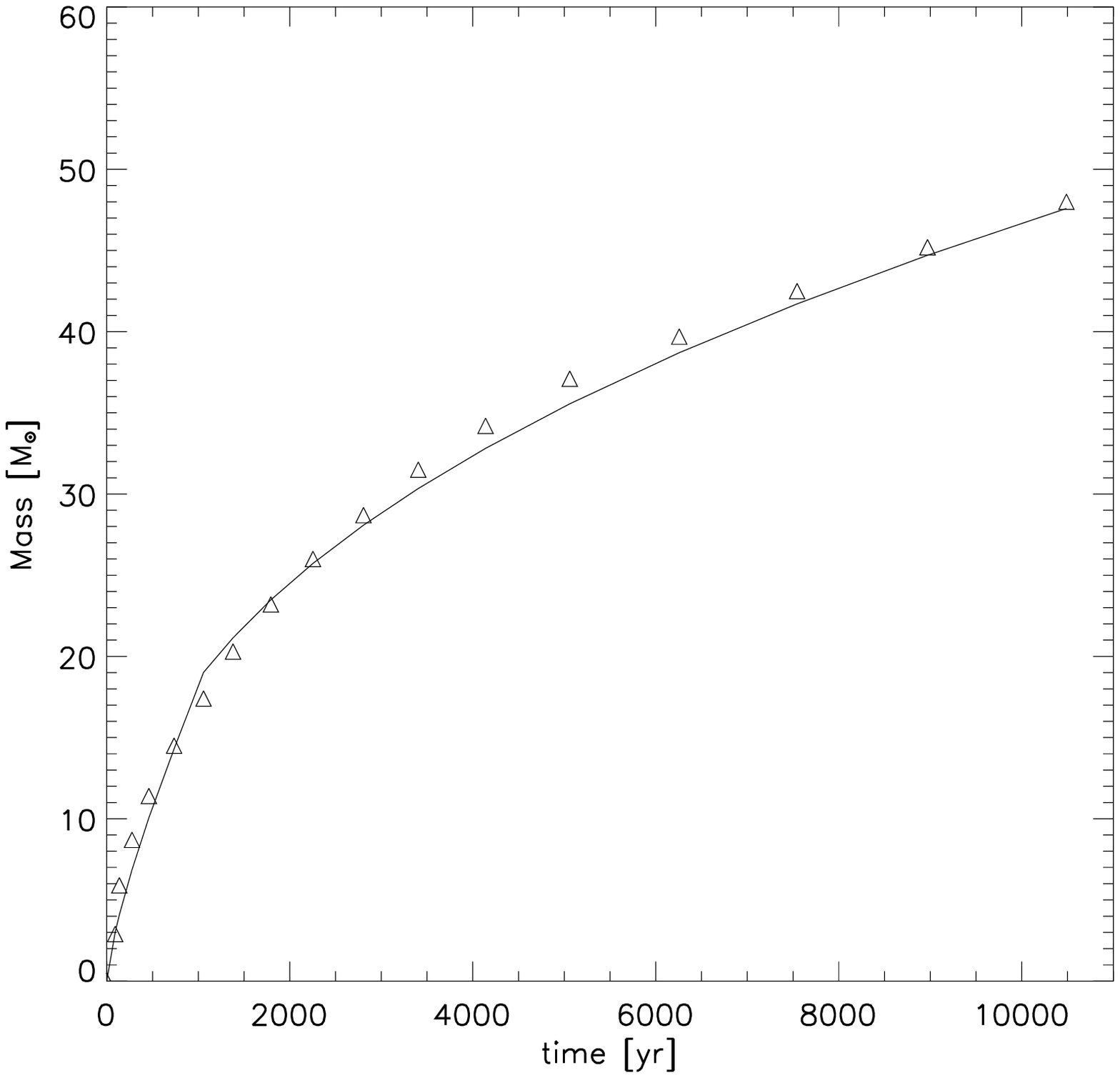,width=6.93cm,height=6.237cm}
\caption{Accretion onto a primordial protostar.  The morphology of the
accretion flow is shown in Fig.~1.  {\it Left:} Accretion rate (in
$M_{\odot}$~yr$^{-1}$) vs. time (in yr) after the formation of a molecular
core.  {\it Solid line:} Accretion rates approximated as: $\dot{M}_{\rm
acc}\propto t^{-0.25}$ for $t<10^{3}$~yr, and $\dot{M}_{\rm acc}\propto
t^{-0.6}$ for $t>10^{3}$~yr.  The early 3D evolution does not follow the
simple power-law temporal behavior found in 1D simulations with very high
resolution (Omukai and Nishi, 1998).  {\it Right:} Mass of the central core
(in $M_{\odot}$) vs. time.  {\it Solid lines:} Accretion history
approximated as: $M_{\ast}\propto t^{0.75}$ for $t<10^{3}$~yr, and
$M_{\ast}\propto t^{0.4}$ for $t>10^{3}$~yr.  Using this analytical
approximation, we extrapolate that the protostellar mass grows to $\sim 120
M_{\odot}$ after $\sim 10^{5}$~yr, and to $\sim 500 M_{\odot}$ after $\sim
3\times 10^{6}$~yr, the lifetime of a very massive star.  }
\end{figure}

In Figure~3, we show the accretion history.  Our time-dependent,
3-dimensional simulation provides the accretion flow around a
primordial protostar and shows how the molecular core grows in mass over
the first $\sim 10^{4}$~yr after its formation. The accretion rate is
initially very high, $\dot{M}_{\rm acc}\sim 0.1 M_{\odot}$~yr$^{-1}$, and
subsequently declines with time in a complex fashion.  Expressed in terms
of the sound speed in Population~III pre-stellar cores, the initial rate
is: $\dot{M}_{\rm acc}\sim 25 c_s^3/G$.  Initial accretion rates of a few
tens times $c_s^3/G $ are commonly encountered in simulations of
present-day star formation (Larson, 2003). We thus find that Population~III
stars reproduce this generic behavior.  The mass of the molecular core,
taken as an estimate for the protostellar mass, grows approximately as
\begin{equation}
M_{\ast}= \int \dot{M}_{\rm acc}{\rm d}t \approx 20 M_{\odot}
\left(\frac{t}{10^{3}\mbox{\ yr}}\right)^{0.75}
\end{equation}
for $t<10^{3}$~yr, and as
\begin{equation}
M_{\ast}= \int \dot{M}_{\rm acc}{\rm d}t \approx 20 M_{\odot}
\left(\frac{t}{10^{3}\mbox{\ yr}}\right)^{0.4}
\end{equation}
afterwards. The early-time accretion behavior is similar
to the power-law scalings found in 1D, spherically symmetric
simulations (Omukai and Nishi, 1998; Ripamonti et al., 2002).
The complex accretion history encountered in our simulation,
however, cannot be accommodated with a single power law. Such
a multiple power-law behavior has previously been inferred 
(Abel et al., 2002; Omukai and Palla, 2003). This earlier inference,
though, was indirect, in that it did not follow the actual accretion
flow, but instead estimated the accretion from the instantaneous
density and velocity profiles at the termination of the 3D
simulation of Abel et al. (2002).
Expressed as a function of stellar mass, our accretion rate scales as
$\dot{M}_{\rm acc}\propto M_{\ast}^{-0.35}$ for $t<10^{3}$~yr, and
$\dot{M}_{\rm acc}\propto M_{\ast}^{-1.5}$ afterwards.
The early behavior is similar to the accretion law 
predicted by Tan and McKee (2003): $\dot{M}_{\rm acc}\propto M_{\ast}^{-0.4}$.
In summary, our simulated accretion history at early times is consistent with
previous estimates, based on 0D and 1D work, but deviates significantly
at later times, at $t > 10^{3}$~yr. Furthermore, the accretion flow 
is not a simple scale-free process when followed over a sufficiently
long time. The self-similarity of the flow is broken because
the presence of three-body reactions introduces a preferred
timescale (see equ. (2)): $t_{\mbox{\scriptsize 3-body}}\sim
10^{3}$~yr.

Our simulations do not take into account any feedback effects from the
proto-star on the accretion flow. To ascertain when these are likely to
intervene, we compare our accretion history with the critical threshold
derived by Omukai and Palla (2003).  We find that as early as $\sim
3000$~yr after the accretion started, the accretion rate drops below
$\dot{M}_{\rm crit}$, at which point the stellar mass is $M_{\ast}\sim 30
M_{\odot}$. The accretion rate becomes sub-critical early on, while the
protostar still undergoes KH contraction. Thus, according to the model of
Omukai and Palla (2003), the star can settle onto the MS without
experiencing a strong continuum-driven wind, and accretion can continue
during the hydrogen-burning stage.  A conservative upper limit for the
final mass of the star is then, $M_{\ast}(t=3\times 10^{6}{\rm yr})\sim 500
M_{\odot}$, assuming that accretion cannot go on for longer than the total
lifetime of a very massive star, which is almost independent of stellar
mass (e.g., Bond et al., 1984; Bromm et al., 2001).

\section{Conclusions}

We have presented the first three-dimensional simulation of Population~III
star formation that followed the accretion flow for $\sim 10^{4}$~yr,
allowing to better estimate the final stellar mass. The earlier collapse
phase prior to the formation of the initial hydrostatic core, gives results
that are similar to previous work (Omukai and Nishi, 1998; Abel et al.,
2002; Ripamonti et al., 2002). In particular, we confirm that the gas is
converted into a fully molecular phase due to three-body reactions within
the central $\sim 100$~AU (comprising an initial mass of $\gtrsim 1
M_{\odot}$). When extrapolating the accretion to $\sim 3\times 10^6$~yr,
the lifetime of a massive star, we get $M_{\ast}\sim 500 M_{\odot}$.

{\it Can a Population~III star ever reach this asymptotic mass limit?} 
Our hydrodynamic simulation lacks radiative transfer
and cannot address this question, as the answer depends on whether the
accretion from the dust-free envelope is eventually terminated by feedback
from the star (e.g., Omukai and Palla, 2001, 2003; Ripamonti et al., 2002;
Omukai and Inutsuka, 2002; Tan and McKee, 2003).  The standard mechanism by
which accretion may be terminated in metal-rich gas, namely radiation
pressure on dust grains (Wolfire and Cassinelli, 1987), is obviously not
effective for gas with a primordial composition. It has been speculated
that accretion could instead be turned off through the formation of an H~II
region (Omukai and Inutsuka, 2002), or through the radiation pressure
exerted by trapped Ly$\alpha$ photons (Tan and McKee, 2003). The unsolved
problem of whether the accretion process is terminated by radiative
feedback from the protostar, defines the current frontier in
three-dimensional numerical studies of the formation of Population~III
stars.

\bigskip
\noindent
{\bf Acknowledgements}

We thank an anonymous referee for constructive comments that helped improve
the paper.  This work was supported in part by NASA grant NAG 5-13292, and
by NSF grants AST-0071019, AST-0204514.

\end{document}